\documentclass[nofigure]{pasj01}
\Received{$\langle$08-Dec-2020$\rangle$}
\Accepted{$\langle$13-May-2021$\rangle$}
\usepackage{times}
\usepackage[pdftex]{graphicx,color}%,hyperref}

\newcounter{num}
\newcommand{\Rnum}[1]{\setcounter{num}{#1} \Roman{num}}

\begin{document}

\title{Polarization images of accretion flow around supermassive black holes: imprints of toroidal field structure}

\author{Yuh \textsc{Tsunetoe}\altaffilmark{1}$^*$
, Shin \textsc{Mineshige}\altaffilmark{1}, Ken \textsc{Ohsuga}\altaffilmark{2}, Tomohisa \textsc{Kawashima}\altaffilmark{3}, Kazunori \textsc{Akiyama}\altaffilmark{4,5,6}
}

\altaffiltext{1}{Department of astronomy, Kyoto University, Kitashirakawa-Oiwake-cho, Sakyo-ku, Kyoto-shi, Kyoto, 606-8502}
\altaffiltext{2}{Center for Computational Sciences, University of Tsukuba, 1-1-1 Tennodai, Tsukuba-shi, Ibaraki, 305-8577}
\altaffiltext{3}{Institute for Cosmic Ray Research, University of Tokyo, 5-1-5 Kashiwanoha, Kashiwa-shi, Chiba, 277-8582}
\altaffiltext{4}{Massachusetts Institute of Technology, Haystack Observatory, 99 Millstone Road, Westford, MA 01886, USA}
\altaffiltext{5}{Black Hole Initiative, Harvard University, 20 Garden Street, Cambridge, MA 02138, USA}
\altaffiltext{6}{National Astronomical Observatory of Japan, 2-21-1 Osawa, Mitaka-shi, Tokyo, 181-8588}

\email{tsunetoe@kusastro.kyoto-u.ac.jp}

\maketitle

\begin{abstract}

With unprecedented angular resolution, the Event Horizon Telescope (EHT) has opened a new era of black holes.
We have previously calculated the expected polarization images of M87* with the EHT observations in mind.
There, we demonstrated that circular polarization (CP) images, as well as the linear polarization (LP) maps, can convey quite useful information to us, such as the flow structure and magnetic field configuration around the black hole. 
In this paper, we make new predictions for the cases in which disk emission dominates over jet emission, bearing Sgr A* in mind. 
Here we set the proton-to-electron temperature ratio of the disk component to be Tp/Te $\sim$ 2 so as to suppress jet emission relative to emission from accretion flow. 
As a result, we obtain ring-like images and triple-forked images around the black hole for face-on and edge-on cases, respectively. 
We also find significant CP components in the images ($\gtrsim 10\%$ in fraction), both with positive and negative signs, amplified through the Faraday conversion, not sensitively depending on the inclination angles. 
Furthermore, we find a ``separatrix'' in the CP images, across which the sign of CP is reversed and on which the LP flux is brightest, that can be attributed to the helical magnetic field structure in the disk.
These results indicate that future full polarization EHT images are a quite useful tracer of the magnetic field structure. 
We also discuss to what extent we will be able to extract information regarding the magnetic field configurations, under the scattering in the interstellar plasma, in future EHT polarimetric observations of Sgr A*.

\end{abstract}

\section{Introduction}\label{Intro}

In April 2019, Event Horizon Telescope (EHT) observations of the supermassive black hole (SMBH) M87* at the center of the giant elliptical galaxy M87 in Virgo cluster presents a scenario in which radiation from the plasma near a black hole is bent by strong gravity and is seen as a ``photon ring'' (\cite{Bar73,Ta04}) or ``black hole shadow'' (\cite{Hi17,von21,Fa00}), providing a strong observational evidence for general relativity (\cite{EHT19a}).
This marks the opening of the new era of black hole research through direct imaging observations with superb angular resolution.
More abundant data of the same or other sources, including those of polarimetry,  are expected to be obtained in near future and will stimulate not only multi-wavelength observations but also theoretical research.

In \citet{Tsu20}, hereafter Paper \Rnum{1}, we performed general relativistic (GR) polarimetric radiation transfer simulations to calculate the expected polarization image of M87* and found that the black hole shadow can be reproduced by the high black hole spin, and that linear polarization (LP)  vectors are disordered by the strong Faraday rotation near the black hole, while the circular polarization (CP) can be amplified by the Faraday conversion in the well-ordered magnetic field in the jet base.
Surprisingly, the CP components are rather strong, comparable to the LP components, although the CP should be negligible in the original synchrotron emission. This is due to the significant Faraday conversion that occurs when the rays pass through hot material threaded with well-ordered magnetic fields. In a model with hot disk, by contrast, the CP images are faint and turbulent, since the hot region occupied with chaotic magnetic fields are Faraday thick so that the Faraday conversion cannot be efficient. 
These results imply that it will be feasible to investigate the magnetic structure,  temperature distribution and coupling between proton-electron near the black hole in M87* through the combination of LP and CP in future observations.

It is well known that M87 belongs to the group of low-luminosity active galactic nuclei  (LLAGNs; \cite{Nag00}), which are thought to be a sort of AGNs that run out of plasma fuel to accrete onto the black hole over long time, and that still exhibit various astronomical phenomena such as jets. 
However, there are also many LLAGNs that do not have powerful jets. Our Galactic Center, Sgr A*, is a good example and has the measured bolometric luminosity of $L_{\rm bol} \sim 10^{-(7-9)} L_{\rm Edd}$ where $L_{\rm Edd}$ is the Eddington luminosity, which is much lower than the typical values ($L_{\rm bol} \sim 10^{-(3-5)}L_{\rm Edd}$) for LLAGNs including M87 (\cite{Na98,BM03}). By comparing these LLAGNs with those exhibiting clear jet images, we expect to find the key factors that determine the existence of jets, and the underlying evolutionary scenario of AGNs.

We know that Sgr A* is also another promising target of the EHT for horizon-scale imaging of a black hole with mass of $\approx 4 \times 10^6M_\odot$, distance of $\approx 8~{\rm kpc}$ (\cite{Gi17,G19}), and the largest angle on the sky of known black holes ($r_{\rm S} \equiv 2GM_{\rm BH}/c^2$, which corresponds to the angle of $\approx 10~{\rm \mu as}$ on the sky).
Because of its proximity, Sgr A* has been actively observed at various wavelengths (e.g. \cite{Ai00,BM03,Gh03,Ei05,Da16}). Recent observations are particularly remarkable in many respects. 
Especially, very long baseline interferometry (VLBI) observations in the millimeter and submillimeter bands and GRAVITY at the Very Large Telescope Interferometer (VLTI) in the near-infrared have provided essential information on the central massive black hole and the accretion flow with high resolution of sub-mas ($\lesssim 100r_{\rm S}$) scale (\cite{Bo03,Bo04,Ma06,Do08,Br11,Fi11,Jo15,Bri16,G17}; 2018a; 2020; \cite{Is19}). 
The accretion rate near the black hole has been estimated to be $\sim 10^{-6}-10^{-9} M_\odot{\rm yr}^{-1}$ from interferometric polarimetry (\cite{Bo03,Ma06}; 2007).
Although Sgr A* is a LLAGN that no clear visual evidence for a jet is known, some intermittent flares and the jet-like shape of the SED have been observed (\cite{Ba01,Ge03,Po03,Ec06,Ma08,Ec12,MF13,G18b}). 
Thus the comparison between these two main EHT targets will provide a detailed understanding of the jet formation scenario.

There are several previous studies of polarimetric radiative transfer simulations near the black hole in Sgr A* (e.g.~\cite{BL06a},b; \cite{Sh12}).
\citet{Go16} compared polarization images for magnetically arrested disk (MAD; \cite{NIA03,Tch11}) and standard and normal evolution (SANE; \cite{Na12,Sa13}) models based on general relativistic magneto-hydrodynamics (GRMHD) simulations, considering the scattering effects in interstellar medium. 
The MAD (SANE) is defined by $\phi \sim 15$ ($\sim 1$) with the absolute magnetic flux $\Phi_{\rm BH} = (1/2) \int_\theta \int_\varphi |B^r| {\rm d}A_{\theta \varphi}$ threading a hemisphere of the event horizon and its dimensionless form $\phi \equiv \Phi_{\rm BH}/\sqrt{\dot{M}r_{\rm g}^2c}$ (\cite{Tch11}).
They concluded that models with ordered magnetic fields, such as MADs, are favored by comparison with visibilities from early EHT observation in 2013 (\cite{Jo15}).
\citet{JRD18} surveyed the electron temperature distributions and the mass accretion rates onto the black hole. 
There, the net linear polarization fraction of Sgr A* favored models with the hot disk and low accretion rate rather than ones with significant jet emission.
Most recently \citet{De20} calculated GRMHD modeling the electron thermodynamics and polarimetric radiative transfer for MAD and SANE models. They also favored the MAD models in terms of spectrum and variability in mm to NIR, source sizes at 230~GHz and 86~GHz, and polarization fraction at 230~GHz, while the SANE models are excluded by low polarization flux due to the strong Faraday  rotation and depolarization.

In this study, we calculate polarized radiative transfer of `semi-MAD' models, with the dimensionless magnetic flux of $\phi \sim 10$ (\cite{Na18}), for disk dominant case, whereas in Paper \Rnum{1} we studied the magnetic structure in the base region of M87 jets using jet dominated models.
We use the parameters of Sgr A* and test the validity of the models by comparing it with polarization observations. Further, we will discuss the morphology of the LP and CP images due to the dominant disk through the Faraday effects (rotation and conversion) and will investigate what and to what extent we can constrain through future polarimetries, considering the interstellar scattering effect unique to Sgr A*. 

The plan of this paper is as follows: 
we will explain methods to obtain polarization images in section \ref{methods}. 
Resultant images for three models will be shown in section \ref{results}, and remarkable features on them and physical processes involved there are described there. 
We will discuss possible effects of the interstellar scattering in future observations with EHT in subsection \ref{convolution} and \ref{capture}, and comparison with past observations and with other models with different inclinations and black hole spin in subsection \ref{obs} and in subsection \ref{opposite} and \ref{a05}, respectively.

\begin{table*}[]
\begin{center}
  \begin{tabular}{c|c||ccc|l|l}
    Reference name & $i$ & ${\rm LP}_{230}$ & ${\rm CP}_{230}$ & $\mathrm{RM}_{\sim230}$ [$\mathrm{rad/m^2}$] & Remarks & Figures \\ \hline
    i30 & $30\degree$ & $4.0\%$ & $0.8\%$& $1.4\times10^5$ & Face-on model & \ref{i30}, \ref{scattered}~top \\
    i60 & $60\degree$ & $5.8\%$ & $0.6\%$& $-1.4\times10^5$ & Intermediate model & \ref{i60}, \ref{scattered}~bottom \\
    i90 & $90\degree$ & $2.1\%$ & $-0.8\%$& $-3.2\times10^4$ & Edge-on model & \ref{i90}, \ref{scattered}~center\\	\hline
    i120 & $120\degree$ & $4.6\%$ & $0.4\%$ & $-1.2\times10^5$ & Intermediate from below & -\\
    i150  & $150\degree$ & $2.0\%$ & $-3.3\%$ &  $1.4\times10^6$ & Face-on from below & \ref{below}\\  \hline \hline
    a05-i60  & $60\degree$ & $0.6\%$ & $-1.5\%$ &$-7.2\times10^5$ &  Slow spin case & - \\
    a05-i120  & $120\degree$ & $2.6\%$ & $-0.9\%$ &5.4$\times10^5$ &  Slow spin from below & \ref{a05-i120} \\  \hline          
  \end{tabular}
\end{center}
  \caption{Calculated models and calculated mass accretion rate, $\dot{M}$, total LP fraction, ${\rm LP}=\sqrt{Q^2+U^2}/I$, total CP fraction with sign,${\rm CP}=V/I$, where $(I,Q,U,V)$ are the Stokes parameters, and rotation measure (RM) calculated from 230 \& 235~GHz simulations. In all models we fix the black hole mass and spin to be \textcolor{black}{$M_\mathrm{BH}=4.5\times10^6M_\odot$} and $a=0.9M_{\rm BH}$. In determination of electron temperature, we set $R_{\rm high}=2$ and $R_{\rm low}=1$ in the relation with proton temperature. Only free parameter is inclination angle $i$, for the top five models. We take a parameter set of $a_{\rm BH}=0.5$ and $R_{\rm high}=1$ for the slow-spin model in the bottom row. The mass accretion rate $\dot{M}$ of $4.0\times10^{-10}M_\odot{\rm /yr}$ ($3.5\times10^{-8}M_\odot{\rm /yr}$) for models with $a=0.9M_{\rm BH}$ ($a=0.5M_{\rm BH}$) is a scaling parameter to the 230~GHz observed flux of Sgr A*, \textcolor{black}{$\approx 3{\rm Jy}$}.}
  \label{models}
\end{table*}

\section{Methods}\label{methods}

Numerical procedures of our study are in two parts: 
(1) GRMHD simulations with data taken from Nakamura et al.~(2018), 
(2) polarized GRRT calculations.

\subsection{GRMHD models of LLAGNs with weak jet}

Our code developed in Paper \Rnum{1} performs three dimensional general relativistic ray-tracing in Boyer-Lindquist coordinates, and transfers the full Stokes parameters $(I,Q,U,V)$ along the light rays from object to the screen.
We adopted the distributions of particle density, proton temperature, magnetic field, and velocity field obtained in axi-symmetric GRMHD simulations by Nakamura et al.~(2018) for our code to calculate polarized radiative transfer.
The synchrotron electron temperature $T_{\rm e}$ is determined from the proton temperature $T_{\rm p}$, using the proton-to-electron temperature ratio relationship by plasma-$\beta (\equiv 2P_{\rm gas}/B^2$) and two parameters $R_{\rm high}$ and $R_{\rm low}$ introduced in \citet{Mo16}, 
\begin{equation}
	\frac{T_\mathrm{p}}{T_\mathrm{e}} = R_\mathrm{high}\frac{\beta^2}{1+\beta^2} + R_\mathrm{low}\frac{1}{1+\beta^2},
\end{equation}
as in Paper \Rnum{1}. Since the factors to convert the density and magnetic field from the GRMHD simulation's unit to the cgs unit depend on the black hole mass $M_{\rm BH}$ and mass accretion rate onto the black hole $\dot{M}$, in this study we use a commonly adopted value of $M_{\rm BH}=4.5\times10^6M_\odot$ and the distance of $8.1~{\rm kpc}$ for Sgr A*, and scale $\dot{M}$ to reproduce its total flux of $\sim$ 3~Jy.

\begin{figure*}
	\begin{center}
		\includegraphics[width=16cm]{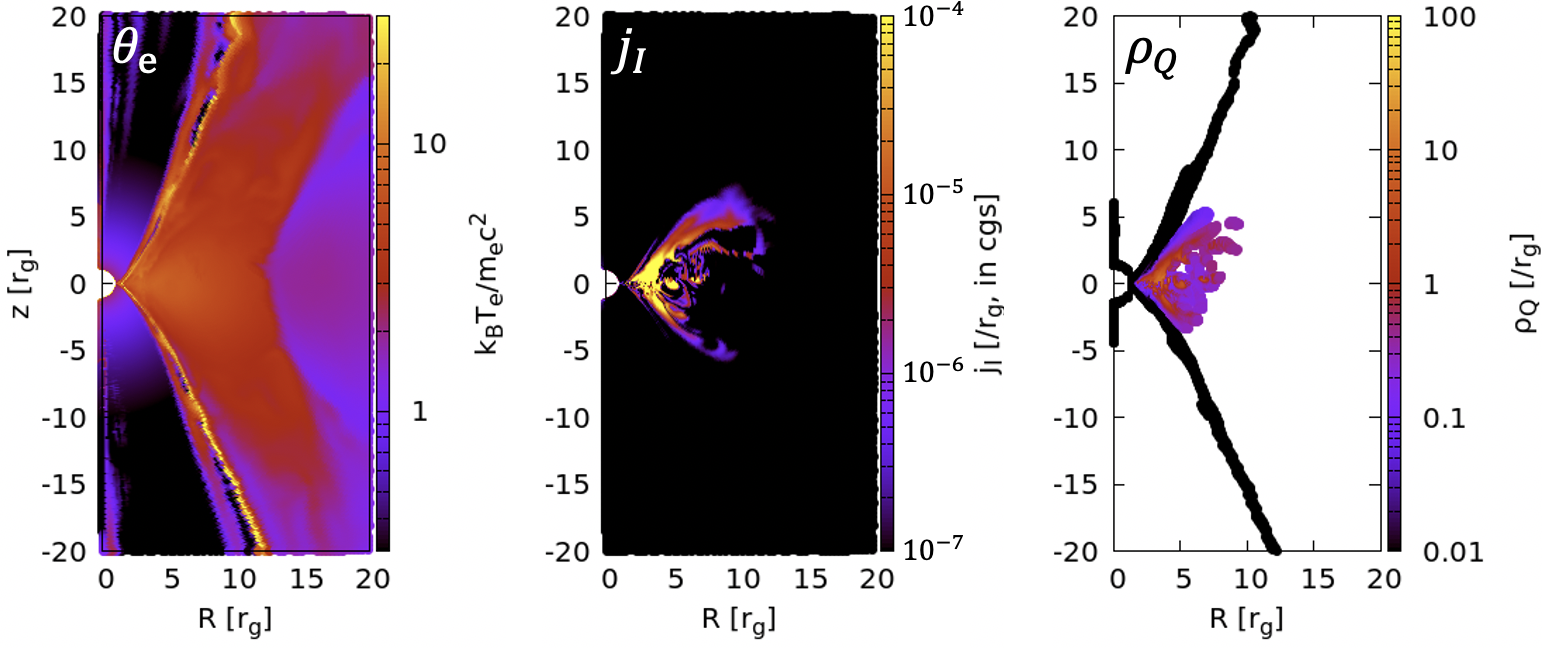}
	\end{center}
	\caption{Two-dimensional distributions of some physical quantities to be used in the polarimetric transfer simulation. Left: map of dimensionless electron temperature $\theta_{\rm e}\equiv \frac{k_{\rm B}T_{\rm e}}{m_{\rm e}c^2}$. Here, $k_{\rm B}$ is the Boltzmann constant, $T_{\rm e}$ is electron temperature, $m_{\rm e}$ is mass of electron, and $c$ is the speed of light. Note that $\theta_{\rm e} =1$ corresponds to $T_{\rm e} \simeq 6\times 10^9 ~{\rm K}$.  Center: map of synchrotron emissivity $j_I$ (in the covariant form, in cgs unit) roughly estimated from electron density, temperature, and magnetic strength. Right: that of the Faraday conversion coefficient $\rho_Q$ (and $\rho_U$, also in the covariant form), but only the region where $|\rho_Q/\rho_V| > 0.01$ ($\rho_V$ is a covariant coefficient of the Faraday rotation) is plotted. See also figures 9 and 10 of Paper \Rnum{1} for more details.}
	\label{Fcon}
\end{figure*}

The purpose of the present study is to study the polarization properties of the LLAGNs which show no visible jets, in the context of disk dominant case for semi-MAD models with strong and ordered magnetic field.
For this reason, we relatively suppress the jet emission (compared with the disk emission) by assigning a low value for $R_{\rm high}$.
The BH spin is $a = 0.9M_{\rm BH}$ (and $0.5M_{\rm BH}$, see subsection \ref{a05},) and the temperature ratios in the disk and jet region are $R_{\rm high} = 2$ and $R_{\rm low} = 1$. Typical electron temperature is $\sim10^{10}~{\rm K}$ in the inner disk region while $\lesssim10^{9}~{\rm K}$ in the jet region, as seen in the left panel of figure \ref{Fcon}. These parameter values are fixed in this study unless noted.
Since full-array EHT data for Sgr A* are not yet available, we do not restrict ourselves to Sgr A* only, but present a typical parameter case of disk dominance and investigate to what extent the results are similar and what differences they exhibit, compared with the previous ones (Paper \Rnum{1}). 
The disk-dominated emission model is also motivated by a possible inefficient injection of the plasma inside the jet funnel in Sgr A*, which is suggested by \citet{BT15}.

\subsection{Polarized radiative transfer calculation}

The method of full-polarimetric radiative transfer calculation is the same as that in Paper \Rnum{1}.
We adopted the viewing angles of the spin axis of $i = 30\degree, 60\degree, 90\degree$, as seen in table \ref{models}, and the observational frequency of 230~GHz.

Here, we note about the numerical setting. In the GRMHD model, some remnant features of the initial torus, which was introduced for the sake of calculation convenience, are left in the outer part of the simulation box. 
In Paper \Rnum{1}, they do not affect the results, since the disk temperature was low and the inclination angles were mostly large ($i=160\degree$) except for some special cases. 
In the present case with hot disk and small inclination angles, however, the calculated images are significantly affected or sometimes dominated by such artificial features, as long as the synchrotron emission, the self-absorption effect or the Faraday effects in the initial torus are not ignored. 
In order to avoid this, we calculate the radiative transfer only in the limited range of $r{\rm sin}\theta \le 10r_{\rm g}$. 
We confirmed that the total flux did not significantly change ($\approx 8\%$ for Model i30) from the case in which the calculation was performed for the whole region $r \le 100r_{\rm g}$, the values of polarization fraction and rotation measure (RM) varied only by a small factor ($\approx 1.2$ for Model i30). 
We thus safely conclude that the omission of the outer zone in the transfer simulation does not change the main outcomes of this paper.

\section{Results}\label{results}

In this section, we will show our results of polarimetric transfer simulations. We will show that the appearance of the LP vector maps and the CP images, as well as the total intensity maps, are sensitive to the inclination angle; that is, they look distinctively for edge-on and face-on observers.

\subsection{Face-on model}

Figure \ref{i30} shows the total intensity map, the LP vector fields, and the CP image for Model i30 from the left to the right, respectively. 
We take a small inclination angle, $i=30\degree$; that is, we see the equatorial plane or accretion disk from nearly the face-on direction. 
The total intensity image is similar to those seen in the previous calculation for M87*  (see, e.g., figure 2 of Paper \Rnum{1}), showing a photon ring of a diameter of $\sim 50~{\rm \mu as}$ formed by light rays passing through close proximity of the black hole horizon, and multiple ring-like features besides the photon ring due to emissions from axi-symmetric intermittent components in the inner disk. 
Especially, the innermost ring-shaped feature corresponds to emission frrom the inner boundary of the disk, i.e.~the innermost stable circular orbit of the particle (ISCO; \cite{Bar72}).
The left side of the multiple rings corresponds to the approaching side of the rotational disk, and therefore is brighter than the other side because of relativistic beaming effects.

The LP image is vertically elongated and much brighter on the left side. 
The contrast between the left and right parts is much more enhanced in the LP image than in the total intensity image. 
Such a different tendency from the total intensity is due to the results of significant Faraday rotation. 
The LP vectors are very well ordered locally, compared with those in Paper \Rnum{1} (see their figure 2). 
The CP image is also characterized by a ring-like feature, as seen in Paper \Rnum{1}. 
However, there is a sign change across the line of $x_{\rm image}$ (Relative RA) $\sim 30~{\rm \mu as}$. We will discuss these features in detail below.

\subsubsection{Sign reversal in the CP image}\label{CPamp}

In equation (24) of Paper \Rnum{1}, we demonstrated that, CP component can be linearly amplified in synchrotron emitting, Faraday thick and optically thin plasma with ordered magnetic field as 
\begin{equation}\label{StokesV}
	V(s) \sim \frac{1}{2}\frac{\rho_Q\rho_V}{\rho_Q^2+\rho_V^2} j_Qs, 
\end{equation} 
where $\rho_Q$ and $\rho_V$ are coefficients of the Faraday conversion and rotation, and $j_Q$ is emissivity of LP component. 
As shown in the central and right panels of figure \ref{Fcon}, the region where synchrotron emission and well-balanced Faraday effects occurs are distributed in the inner disk near the black hole ($\lesssim 5r_{\rm g}$), more widely than in Paper \Rnum{1}. Thus the CP amplification significantly occurs here. 
In a previous work, \citet{Ho09} introduced a combination of Faraday rotation and conversion to model the CP observed in a quasar 3C~279, mentioned it as `rotation-driven conversion'.
Equation (\ref{StokesV}) is an approximate form, ignoring oscillation terms, of the one derived by \citet{De16} (see their appendix C). 

\begin{figure*}
	\begin{center}
		\includegraphics[width=16cm]{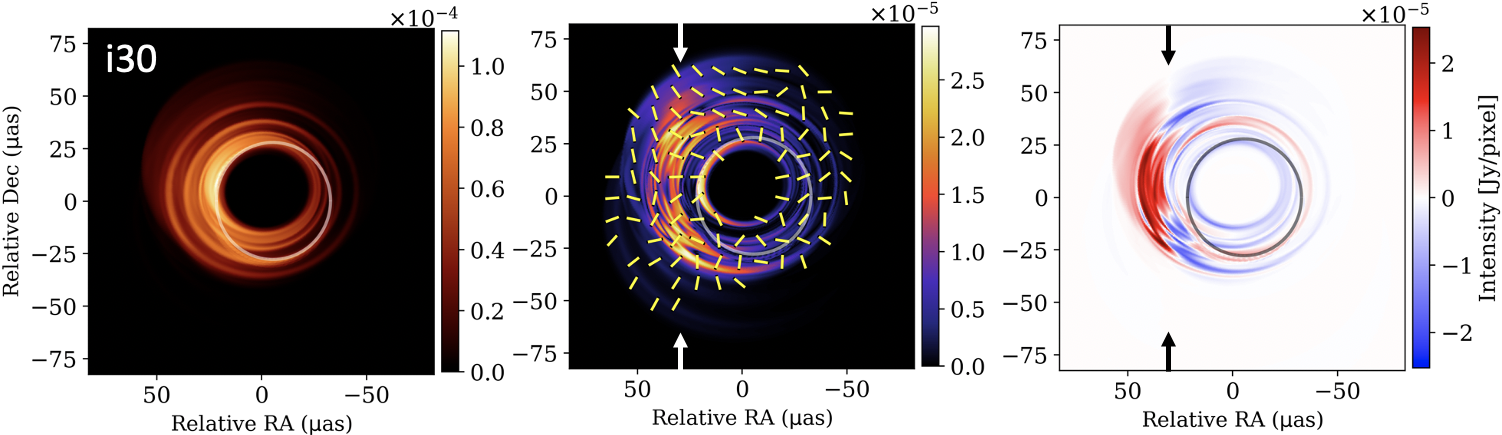}
	\end{center}
	\caption{Polarization images of Model i30. Left: total intensity (Stokes $I$) image. Center: Linear polarization (LP) map with color contour of LP intensity (Stokes $\sqrt{Q^2+U^2}$) and LP vectors in EVPA (electric vector position angle). Right: circular polarization (CP) image with color contour of CP intensity with sign (Stokes $V$). The spin axis of the black hole points upwards in the figures. Solid circles in the images correspond to the photon ring analytically obtained for the BH spin and the inclination angle (\cite{Ta04,Jo13,Wo20,Ka20}). Two arrows in the top and bottom of the LP map and CP image indicate the position of the ``separatrix'' line (i.e., it lies in between the two arrows) described in subsubsection \ref{CPamp} and \ref{separatrix}.}
	\label{i30}
\end{figure*}

\begin{figure*}
	\begin{center}
		\includegraphics[width=8cm]{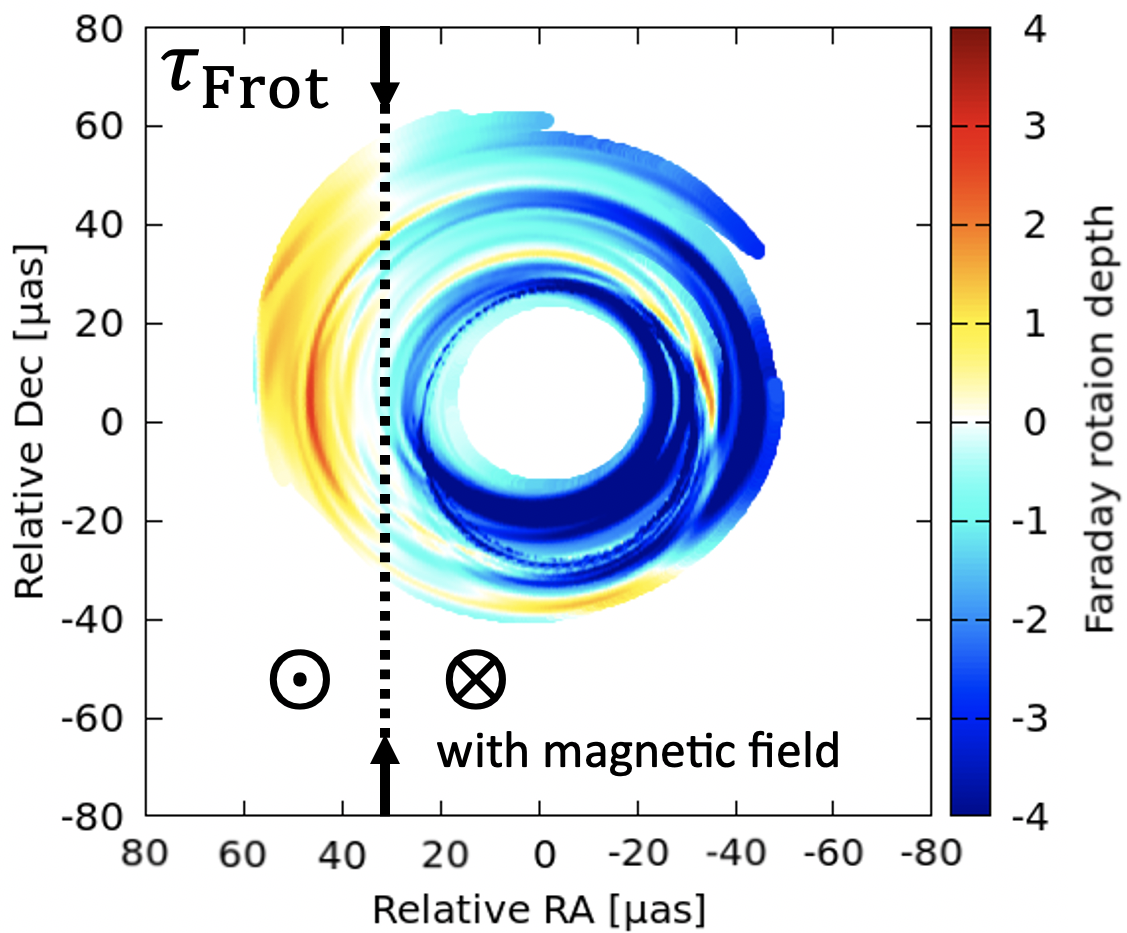}
	\end{center}
	\caption{Image map of Faraday rotation depth $\tau_{\rm Frot} \equiv \int \rho_V {\rm d}\lambda$ for Face-on model. Only light rays with $\tau_{\rm Fcon}\equiv \int\sqrt{{\rho_Q}^2+{\rho_U}^2}{\rm d}\lambda>0.2$ are plotted, focusing on the rays whose CP components can significantly be amplified.}
	\label{tauFrot}
\end{figure*}

Since the CP is amplified through the relation, $V(s) \propto \rho_V \propto {\rm cos}\theta_B$, an important key factor of amplification is $\theta_{\rm B}$, angle between the line of sight and the direction of magnetic field lines. 
This means that the sign (or direction) of the CP amplification depends on whether the line of sight and the direction of magnetic field line are close to parallel ($0 \le \theta_B < \pi/2$) or anti-parallel ($\pi/2 < \theta_B \le \pi$).
In the present model, magnetic field in the amplification region (i.e., the inner disk) is toroidally dominated, 
so that the sign of ${\rm cos}\theta_B$ should be positive (or negative) on the left (right) side of the observational screen. 
Here, we define the depth of the Faraday rotation as $\tau_{\rm Frot} \equiv \int \rho_V {\rm d}\lambda$, where $\lambda$ is an affine parameter of a light path, and illustrate the Faraday rotation map for Model i30 in figure \ref{tauFrot}. 
We find in this figure that the positive and negative sign reversal occurs across the line of $x_{\rm image} \approx 30~{\rm \mu as}$\footnote{The border is not at $x_{\rm image}=0$, because of the parallax (aberration) effects of special relativity by bulk motion of plasma. See appendix \ref{nomotion} for details.}.
In the CP image shown in the right panel of figure \ref{i30}, the CP component changes its sign across the line at $x_{\rm image} \approx 30~{\rm \mu as}$, although there are some exceptional contaminants. 
We can thus concluded that the sign reversal on the CP image traces the direction of Faraday rotation on the screen, and reflects the toroidal magnetic structure in the inner disk.

\subsubsection{The LP flux on the CP separatrix}\label{separatrix}

The LP image in the central panel of figure \ref{i30} has the vertically elongated feature, as mentioned above, at $x_{\rm image} \approx 30~{\rm \mu as}$, where the CP components changes its sign in the right panel. 
Thus, the LP flux tends to be enhanced where the CP flux vanishes.
This brightened LP flux on the ``separatrix'' of the CP component can be easily explained in relation to the above statement about the angle dependence of the Faraday rotation.

As seen in figure \ref{tauFrot}, the Faraday rotation becomes very weak ($\tau_{\rm Frot}\ll 1$) in the case where the toroidal magnetic field becomes perpendicular with the line of sight at $x_{\rm image} \approx 30~{\rm \mu as}$. 
Thus LP components are not Faraday-rotated so that the CP component should vanish around the vertical line of $x_{\rm image} \approx 30~{\rm \mu as}$.
In other regions on the image, the Faraday rotation depths are relatively high ($|\tau_{\rm Frot}| \sim 1$) regardless of their sign, thus the LP vectors are Faraday-depolarized and decline their polarized fractions. 
There, CP components are amplified to positive (or negative) values on the left (right) side of the separatrix as a consequence of the Faraday rotation.
In other words, the sign reversal of the CP components and the large values of the LP component are mutually related; the Faraday depolarization processes are inefficient around the separatrix (where $\theta_B \approx \pi/2$), which results in large LP components and negligible CP components. 
Therefore, these features on the polarimetric images suggest that, we can expect to give a strong constraint to the configuration of magnetic structure, through comparison between linear and circular polarimetry in future observations.

\subsection{Edge-on model}\label{edge-on}

\begin{figure*}
	\begin{center}
		\includegraphics[width=16cm]{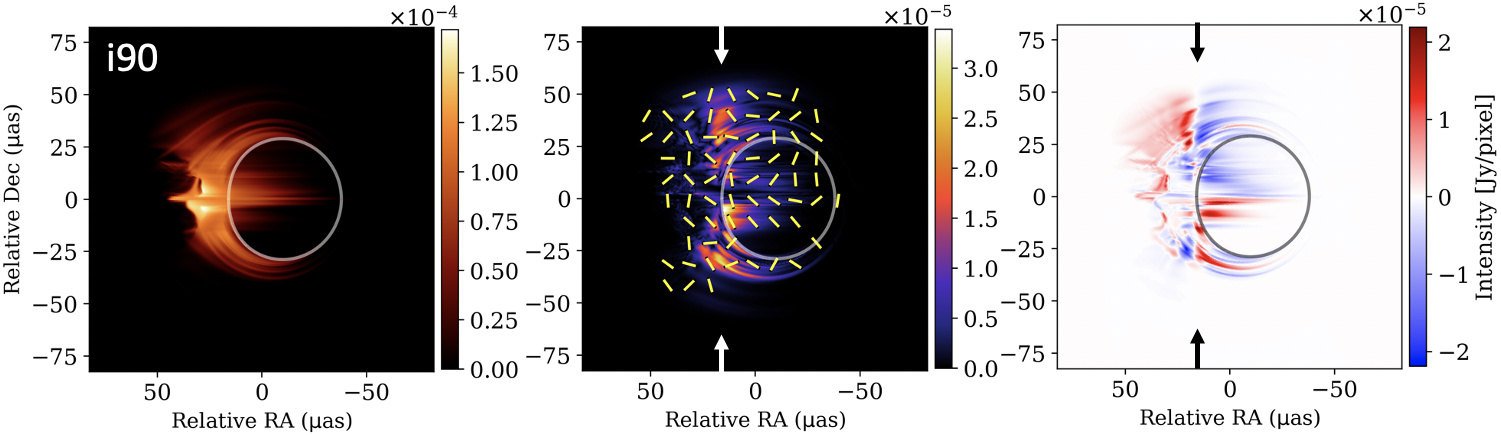}
	\end{center}
	\caption{Same as Fig. \ref{i30} but for Model i90.}
	\label{i90}
\end{figure*}

Figures \ref{i90} illustrate the results of Model i90 with a $i=90\degree$ inclination, i.e., the disk is seen from an edge-on observer. 
Total intensity image in the left panel shows a triple-forked shape. 
Note that the image is brighter on the left side, while the ring-like part is fainter. 
This is because the optically thick disk with high electron temperature and density appears on the image near the equatorial plane, when seen from the edge-on direction.
Further, the relativistic beaming effect due to the Keplerian rotation gives an asymmetric feature of the ``crescent''-shaped shadow, in addition to the foreground disk component. 
Similar fork-like features due to the asymmetric shadow and the foreground disk can also be  seen in edge-on cases in previous works (\cite{Mo14,Pu18,An20}). 

The LP components in the central panel are also brighter on the left side, but fainter at the `root' of the triple-forked, which is the brightest in the total intensity image. 
The LP vectors have different orientations in the upper and lower branches of the triple-forked image, indicating an asymmetry in the Faraday rotation effects. 
The CP image in the right panel is bright tracing the fork and contains the sign reversal at around $x_{\rm image} \approx 15~{\rm \mu as}$. 
In addition, it is interesting to note another sign reversal occurring in the upper and lower portions across the line of $y_{\rm image} = 0$. 
This latter sign reversal in vertical direction reflects the fact that the direction of magnetic field is opposite above and below the equatorial plane, because the magnetic field lines are frame-dragged by the rotational motion of the disk.
Comparing LP and CP images, we also see the features due to the helical field structure,  described in the previous section for the face-on model. In both of the above and below the equatorial plane, the bright LP flux region is vertically elongated at around $x_{\rm image} \approx 15~{\rm \mu as}$, where the signs of the CP components change.

\begin{figure*}
	\begin{center}
		\includegraphics[width=16cm]{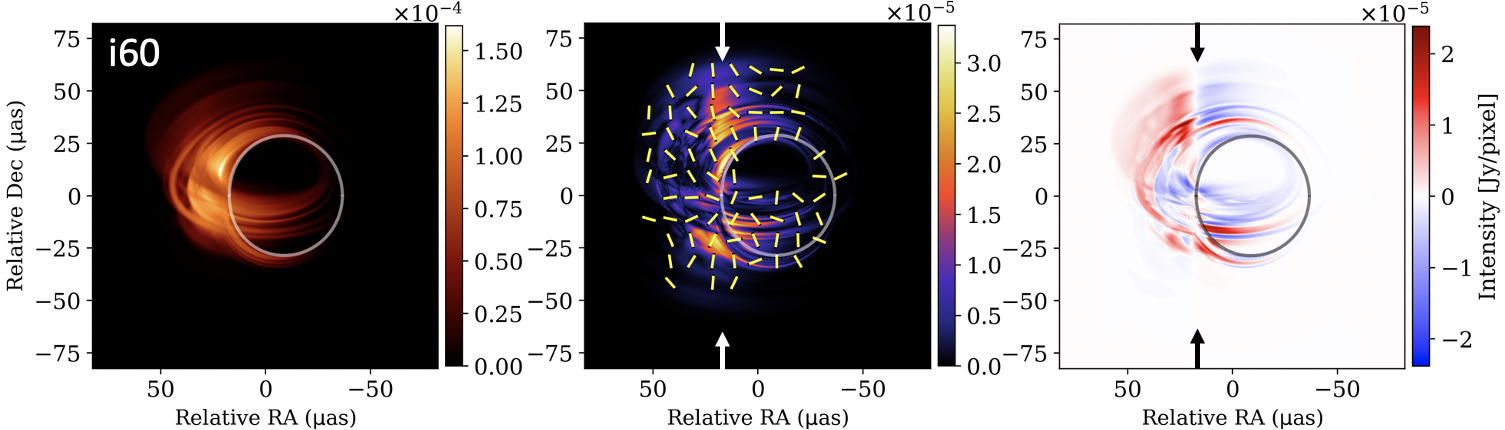}
	\end{center}
	\caption{Same as Fig. \ref{i30} but for Model i60.}
	\label{i60}
\end{figure*}

\subsection{Intermediate model}

Figure \ref{i60} shows the results of the cases with $i=60\degree$, which is often used as a `conservative' value for Sgr A*. 
In short, the resultant images and polarization features are just in between the nearly face-on case (Figure \ref{i30}) and the edge-on case (Figure \ref{i90}).
We can see in the total intensity map both of a ring shape and a dim disk feature. This image is consistent with their images for SANE models by a previous work by Chael et al.~(2018), in terms of composition of the photon ring and foreground disk relativistically beamed in the left side on the image.

The LP intensity is brighter on the left side, but fainter near the base of the fork-like, as in the edge-on model. The polarization vectors are well ordered in the upper half of the image, but are relatively disordered in the lower half. 
This LP map can be clearly compared with those in \citet{JRD18}, since it has the base of an axi-symmetric GRMHD model and a moderate inclination in common with theirs. 
Our LP map of Model i60 resembles the bottom right panel of their figure 2 in the ordered upper-half of the image and the disordered lower-half, as described above, in addition to the Stokes $I$ image consisting of the photon ring and the disk. 
Note that their LP maps comprise of the color contour of Stokes $I$ and the LP vectors.
Next we focus on their models with total flux of $F_\nu = 3 {\rm Jy}$ and a temperature parameter of $\mu =2$ to compare the model parameters, noting that the latter $\mu$ corresponds with our $R_{\rm high}$. 
They found accretion rates of $\dot{M} \simeq 4\times10^{-10}-5\times10^9 M_\odot{\rm yr^{-1}}$, image-averaged Faraday rotation depths of $\langle\tau_{\rm Frot}\rangle \simeq 1-10$, and the net linear polarization fractions of $\pi \le 5\%$. 
All of these ranges of values are consistent with ours for Model i60.

The CP tends to be respectively positive and negative on the left and right side of  the border at $x_{\rm image} \approx 15~{\rm \mu as}$, similarly to the above models, although there exist minor exceptional components. 
These `contaminants' appear due to the violation of the conditions for the CP amplifications (see subsubsection \ref{CPamp}); that is, due to large optical depth and/or significant turbulent field configurations.

In the central panels of figures \ref{i30}, \ref{i90}, and \ref{i60}, the LP flux is weaker on the photon ring than otherwise. 
There are two main reasons for these depolarizations: 
(1) Light rays are strongly bundled around the photon ring so that they become Faraday thick ($\tau_{\rm Frot} \sim 5-10$) at 230~GHz because of long path lengths, while they are moderately thin ($\tau_{\rm Frot} \sim 1$) other area on the images (see figure \ref{tauFrot} for Model i30). Therefore, the LP components are strongly depolarized on the photon ring. 
(2) The rays emitted from distinct areas with different magnetic field configurations (producing different polarization components) are mixed together due to the photon sphere effects, leading to the cancellation of LP components. We wish to remind that the polarization angles vary between the direct ($n=0$; $n$ is the number of half-orbits) and indirect ($n=1,2,3,...$) images (see \cite{Him20}). 
In fact, \citet{Ji21} reported the depolarization of the photon ring by a factor of $\sim 2$ by subtracting the LP flux from the total flux in each pixel.
Note that the depolarization of the LP flux can occur due to the second effect, even if the plasma is Faraday thin and the magnetic fields are ordered (which may occur, for example, at higher frequencies).

\begin{figure*}
	\begin{center}
		\includegraphics[width=16cm]{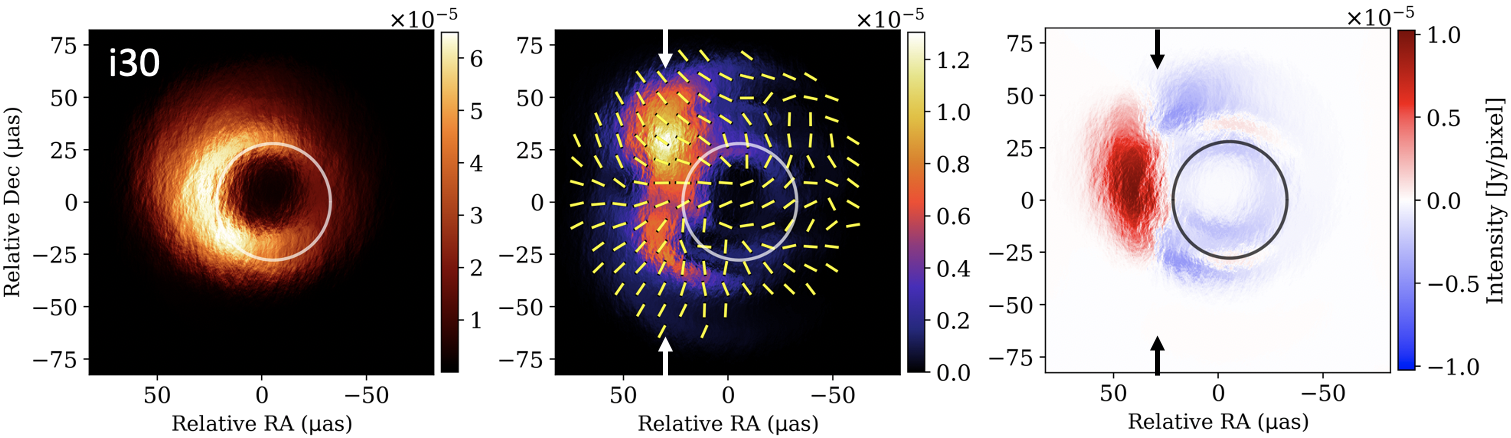}
		\includegraphics[width=16cm]{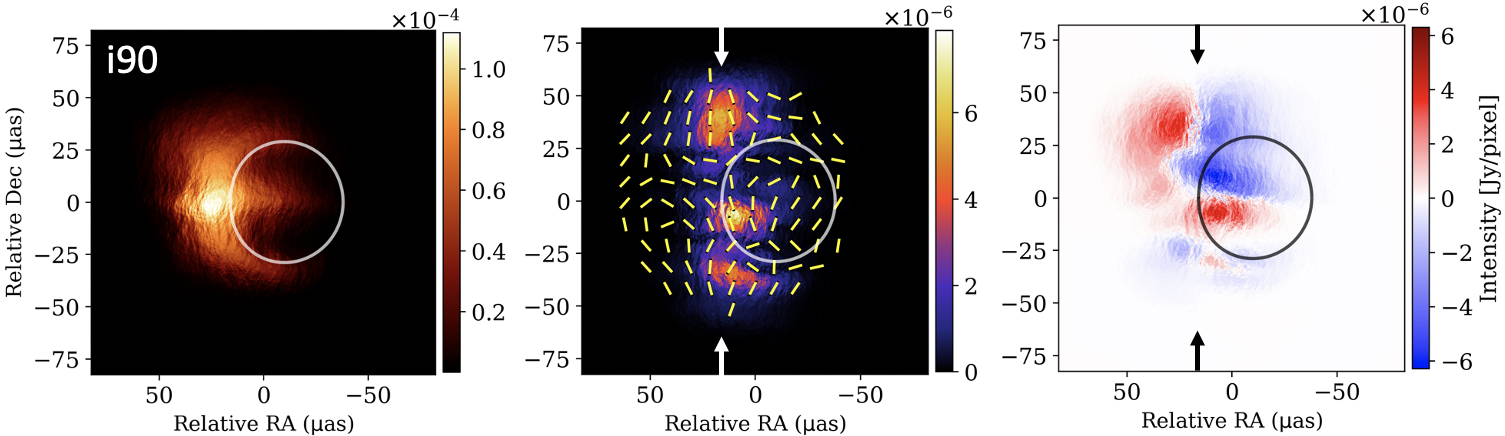}
		\includegraphics[width=16cm]{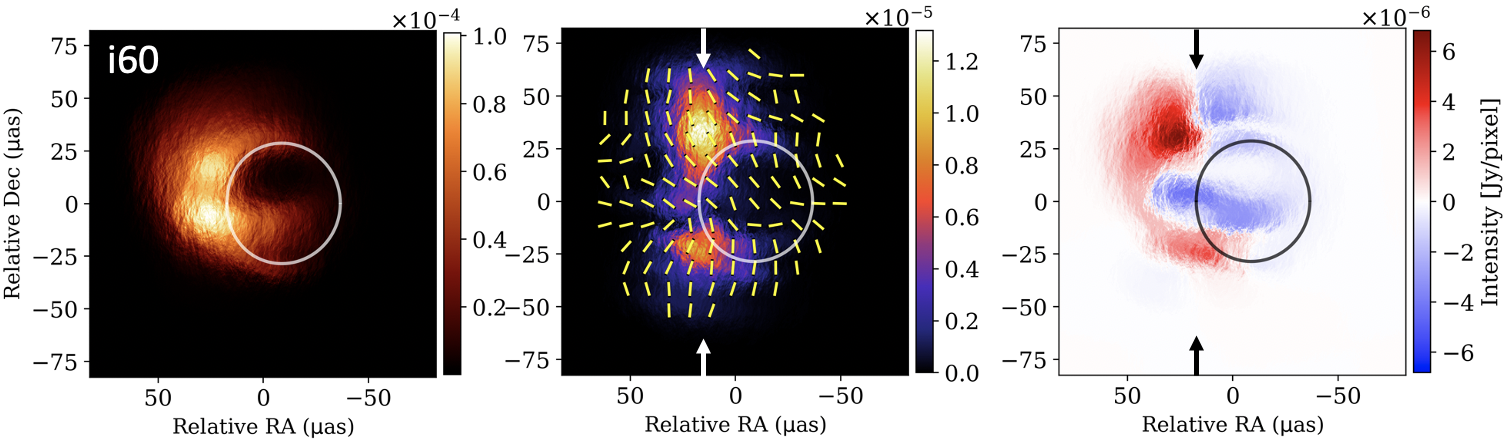}
	\end{center}
	\caption{Top to bottom: scatttered images of Face-on (i30), Edge-on (i90), and Intermediate (i60) model, respectively. Left to Right: the total intensity images, the LP maps, and the CP images. Two arrows on the top and bottom of the LP maps and the CP images correspond to those in the raw images in Fig. \ref{i30}, \ref{i90}, and \ref{i60}.
	}
	\label{scattered}
\end{figure*}

\section{Discussion}\label{discussion}

\subsection{Scattering effects in interstellar medium}\label{convolution}

In the previous section, we presented the `raw' polarization images from radiative transfer calculations and explained the physical processes producing the notable features. 
However, in actual observations of Sgr A* it is known that the tenuous interstellar plasma  along the light of sights scatters its radio waves from the object, affecting its appearance in radio wavelengths (\cite{Da76}). 
The scattering effects on Sgr A* is in the regime of the strong scattering (e.g. \cite{Na92}), where their image distortion is well described with two subdominant effects, diffractive and refractive scattering (\cite{BN85,NG89,GN89}). Diffractive scattering will cause the angular broadening, described as the blurring convolution of the image with a scattering kernel (e.g. \cite{Fi14}). On the other hand, refractive scattering will create stochastic compact substructures on the image (\cite{Jo15,JN16,Ps18,Jo18}). 
In the present work, we implement these scattering effects into our polarimetric images with two software libraries {\tt SMILI} (\cite{Ak17a}, b) and {\tt eht-imaging} (\cite{Ch16}; 2018).

\begin{figure*}
	\begin{center}
		\includegraphics[width=16cm]{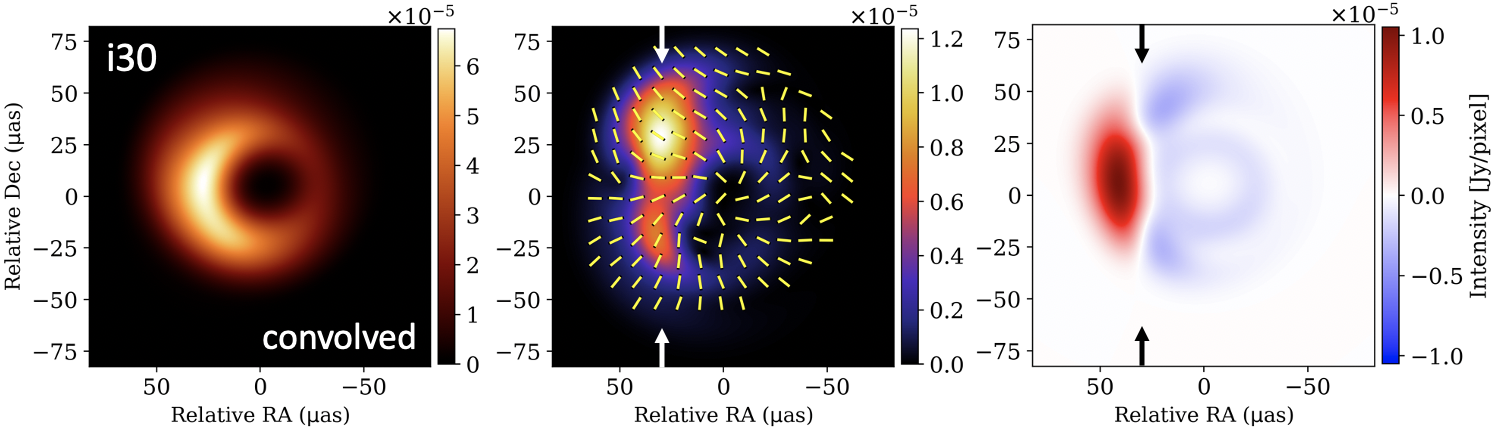}
		\includegraphics[width=16cm]{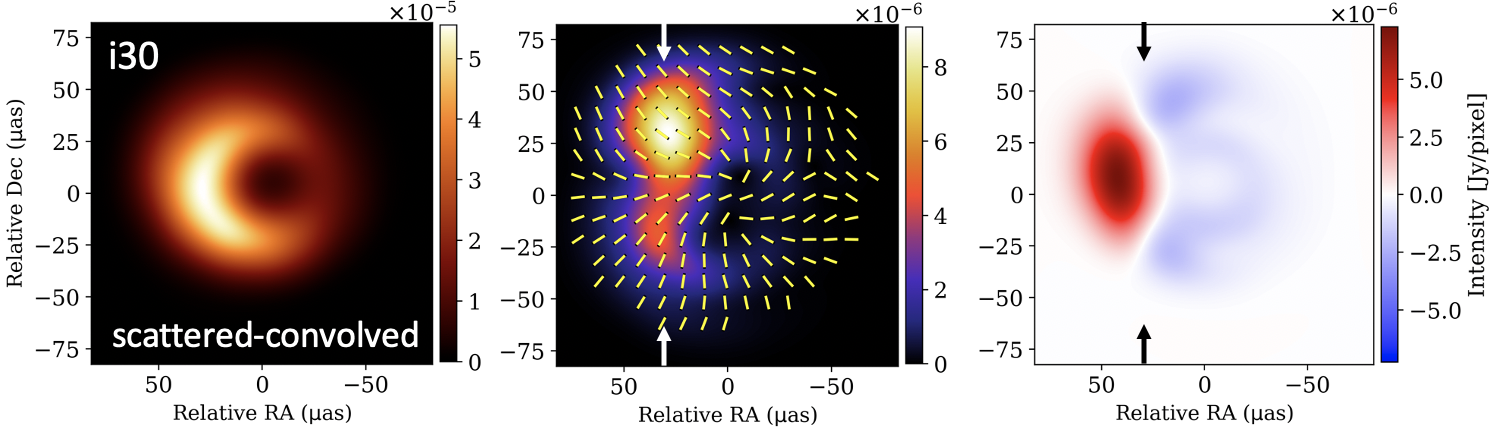}
		\includegraphics[width=16cm]{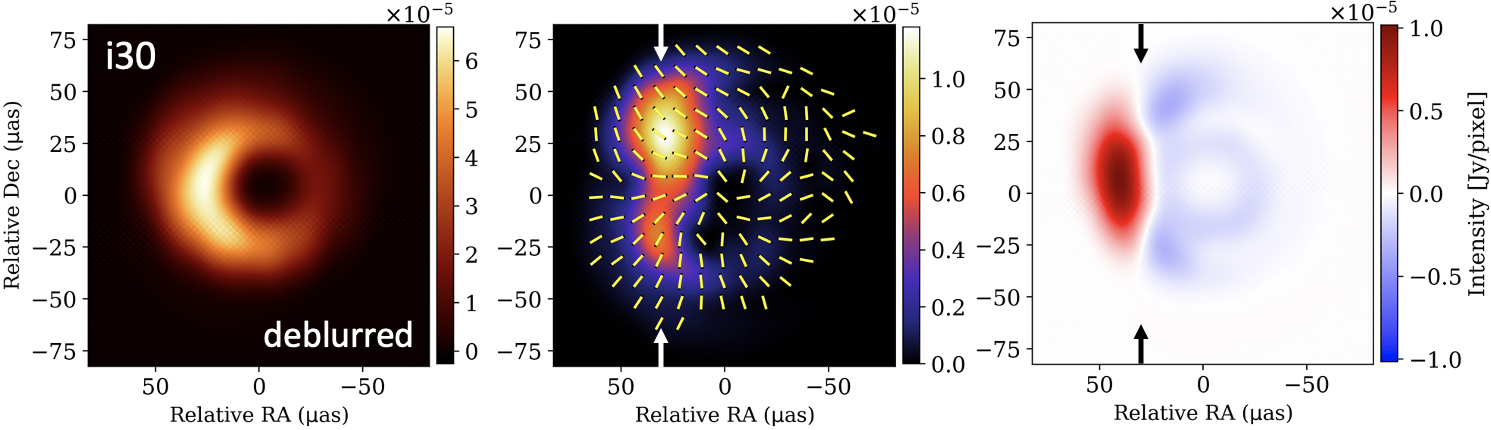}
	\end{center}
	\caption{Top: unreal images for Model i30 convolved with Gaussian beam of $17~{\rm \mu as}$ without the scattering effects. Middle: mock-observational images scattered and Gaussian-convolved. Bottom: scattered, convolved, and deblurred images. Left to Right: the total intensity images, the LP maps, and the CP images. Two arrows on the top and bottom of the LP maps and the CP images correspond to those in the raw and scattered images.
	}
	\label{convo}
\end{figure*}

\subsubsection{Ring-like versus fork-like features in Stokes $I$ images}

In figure \ref{scattered}, we show scattered images of all three models shown in section 3.  
The images at $230~{\rm GHz}$ are blurred with the diffractive kernel of $\sim 10 {\rm \mu as}$. 
In addition, the refractive effects introduce substructures into the images, although they do not affect the features discussed here.
The scattered images in the left column, of total intensity, give the feature referred below as ring-like and fork-like for the face-on model i30 and edge-on model i90, respectively. In the intermediate model i60, the image has the intermediate feature of the small shadow in the upper half and the foreground disk in the lower half. These results imply that we may constrain the inclination angle for Sgr A* through morphology of the  image, with the ring extraction from the image (\cite{Ps15,Ku18,EHT19d}), although their dependence on the time-variability and three-dimensional features in the flaring and non-flaring states should be checked in future works.

\subsubsection{Morphology of linear and circular polarimetric images}

As shown in the images in the right column of figures \ref{scattered}, the scattered CP images still retain the characteristic separatrix features; 
the CP component is positive (or negative) in the left (right) parts. 
Rather, these features in the scattered images are simpler and even more enhanced than in the raw CP images (see figures \ref{i30}, \ref{i90}, and \ref{i60}), because the additional minor features tend to be erased by the diffractive blurring effect. 
The scattered LP images in the central column of figures \ref{scattered} clearly exhibit the brightened-on-separatrix features explained in subsubsection \ref{separatrix}. 
Namely, we see that the LP component flux reaches its maximum, while the sign of the CP component is reversed. 
We thus conclude that the information of magnetic field structures around the black hole will be mostly preserved even with the scattering effects. 
If we find the separatrix feature, in particular, that will be a good indicator of toroidal-dominated field structures.
In \citet{Go16}, the LP and CP images (also including the scattering effects) for two SANE models shows similar features (see their figure 7). Especially, their {\tt SANE\_dipole-jet} model, with inclination of $i=126\degree$ and spin of $a=0.92M_{\rm BH}$, gives the CP image with the vertical separatrix at $x_{\rm image} \approx 15-20 ~{\rm \mu as}$ and the LP component brightened along the separatrix.

\begin{figure*}
	\begin{center}
		\includegraphics[width=16cm]{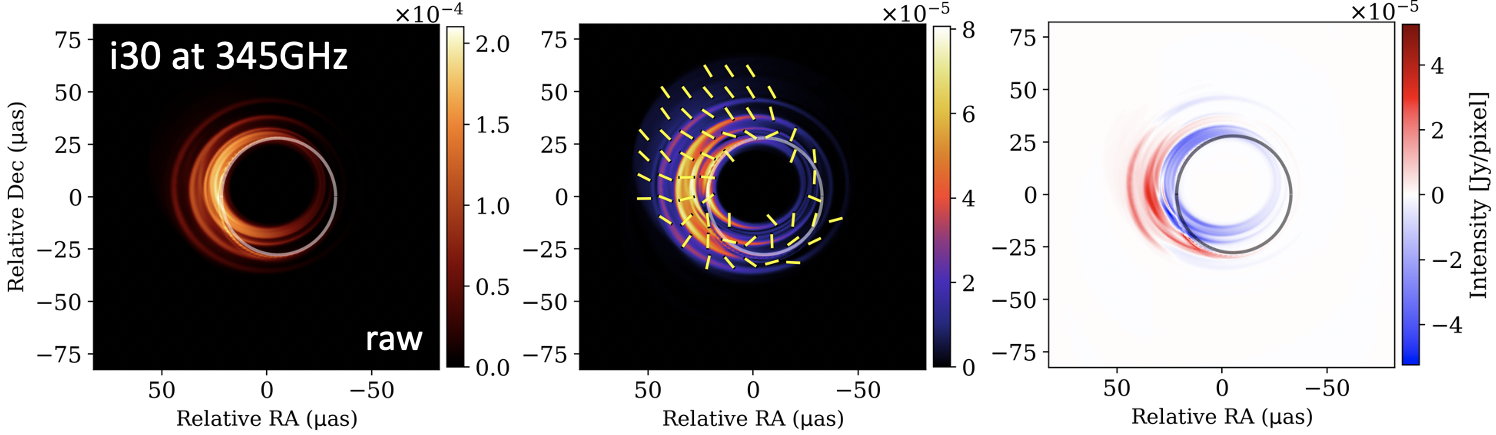}
		\includegraphics[width=16cm]{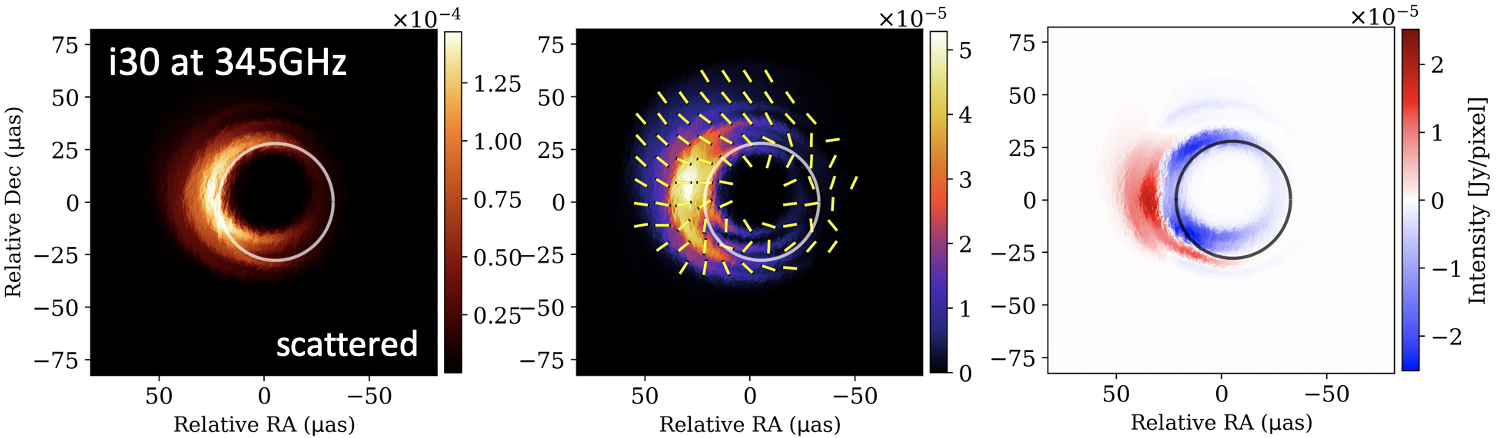}
		\includegraphics[width=16cm]{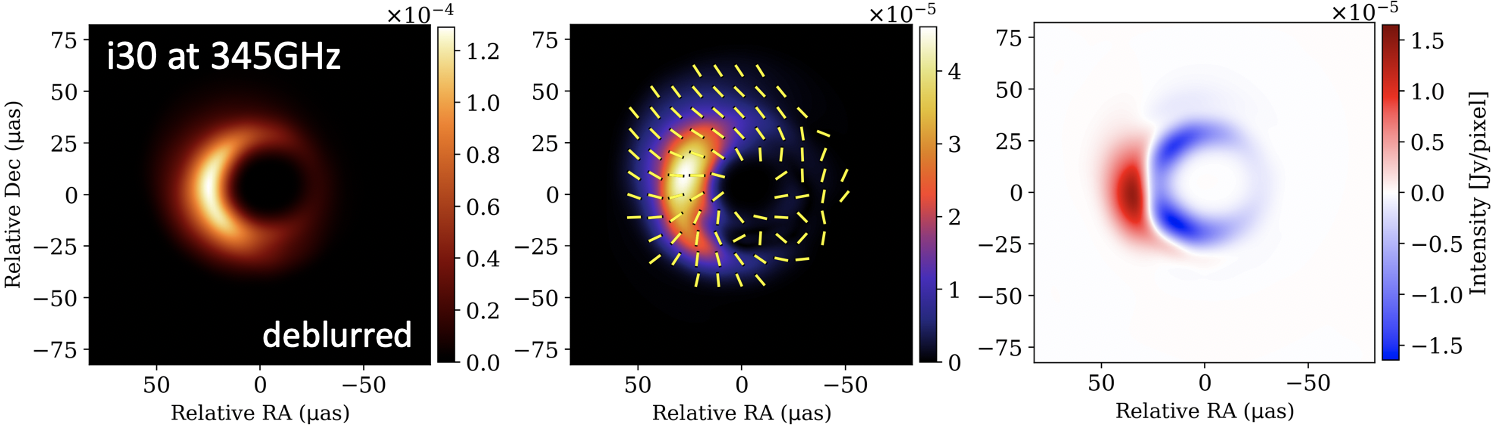}
	\end{center}
	\caption{Top to bottom: raw, scattered, and observational (deblurred) images at 345~GHz for Model i30, respectively. Left to Right: the total intensity images, the LP maps, and the CP images.
	}
	\label{345im}
\end{figure*}

\subsection{Capture of the polarimetric features in future observations}\label{capture}

In the last subsection, we demonstrated that the major features in our raw polarization images (so intrinsic to plasma and radiative physics in the object Sgr A*) are resilient to scattering effects. 
Here, we will discuss the feasibilities for us to capture the features at the angular resolution of the EHT, by blurring the images of our representative Model i30. 
Here, we adopt a blurring Gaussian beam size of $17~{\rm \mu as}$, a resolution expected in recent and future EHT observations for Sgr A* at 230~GHz (\cite{EHT19b}). 

Here, we show three types of processed images to demonstrate how scattering and its popular mitigation work on our simulation images. 
First, we give beam-convolved unscattered images (i.e. figure 2 images convolved with the beam) in the top of Figure 7, for an ideal observed image without scattering effects. 
Second, we yield beam-convolved scattered images to show images with full observing effects (i.e. we convolve the scattered images in the top of figure \ref{scattered}) in the middle of figure \ref{convo}. 
We finally show beam-convolved images with the mitigation of diffractive scattering effects by de-blurring the scattered images with the diffractive scattering kernel (\cite{Fi14,Jo15}). 

Comparing the top and middle row of figure \ref{convo}, we see at first glance that the two results make no large difference because the kernel of diffractive scattering is comparable or less than the observational resolution, and also because the substracture of refractive scattering is not dominant and washed out at the EHT's resolution. 
Carefully watching them, we can point out that there exist some deviations in small scale, such as in the `separatrix' on the CP image, but subsequently find out that these discrepancy on the observation are resolved through the deblurring process in the bottom row of figure \ref{convo}. 
The good agreement between in the top and bottom row implies that even the refractive effect, which is not invertible with deblurring (\cite{JG15}), is faint and compact enough to significantly affect the polarimetric images in the observations at present, although it may be problematic and require the direct modeling of refractive phase screens (\cite{Jo16}) in future observations with higher resolutions.

\subsection{Polarimetric images at 345~GHz}\label{345}

With future observations at higher frequency in mind, we also present polarimetric images at 345~GHz for Model i30 in figure \ref{345im}. 
The raw images in the top row show the more compact emissions near the black hole in the left, more ordered LP vectors in the center, and weaker CP components in the right than the raw images at 230~GHz in figure \ref{i30}, because both of the synchrotron self-absorption and the Faraday effects (rotation and conversion) are weaker at shorter wavelengths. 
That is, the photosphere of synchrotron emissions locates near the black hole and the LP vectors can keep the original information about the magnetic fields in the emission region, although the amplified CP components with the separatrix of their sign can be seen to some extent. 

We can see that these polarimetric features survive from the interstellar scattering in the images in the central row, because both diffractive and refractive scattering effects are also weaker at shorter wavelength (e.g. blurring kernel size $\propto \lambda^2$ at submillimeters; \cite{Jo18}). Further, since the resolution is improved at short wavelengths, we can expect that the directly-interpretable images will be obtained at 345~GHz, as in the observational (deblurred after scattered and convolved with beam of $11~{\rm \mu as}$) in the bottom row.

\subsection{Comparison with existing observations}\label{obs}

Here we test the validity of the models by comparing them with existing observations. 
There have been many polarimetric observation of Sgr A* at 230~GHz mainly since 00s (\cite{Ai00,Bo03}). 
During about twenty years, the LP fraction of $\sim10\%$ (\cite{Bo18}) and CP fraction of $\lesssim 1\%$ (\cite{Mu12}) have been detected with some long-term variabilities. 
On the basis of these observations, a previous study \citet{De20} imposed constraints of $LP = 2-8\%$ and $|CP| = 0.5-2\%$ on their radiative transfer models. 
Checked by their criteria for polarization fractions, our three models of i30, i60 and i90 pass the tests for both of the LP and CP, as seen in table \ref{models}. 
In addition, Sgr A* at around 230~GHz has shown the RM of about $10^5-10^6{\rm rad/m^2}$ (\cite{Ma07}). This RM has been thought to originate in the extended accretion flow up to the scale of $\sim 10^5r_{\rm g}$ (`external' RM), while RMs obtained from our models in table \ref{models} are due to the Faraday rotation within $\sim10r_{\rm g}$ near the emission region (`internal' RM). Thus we use the observational RM value as a rough upper limit estimated in order of magnitudes. In this sense, all the three models also pass the limitation test, and are consistent with the RM observations. 
In this way, we demonstrated that our three models can reproduce the features in polarimetric observations at 230~GHz, whereas some models additionally introduced below violate these constraints.

\begin{figure*}
	\begin{center}
		\includegraphics[width=16cm]{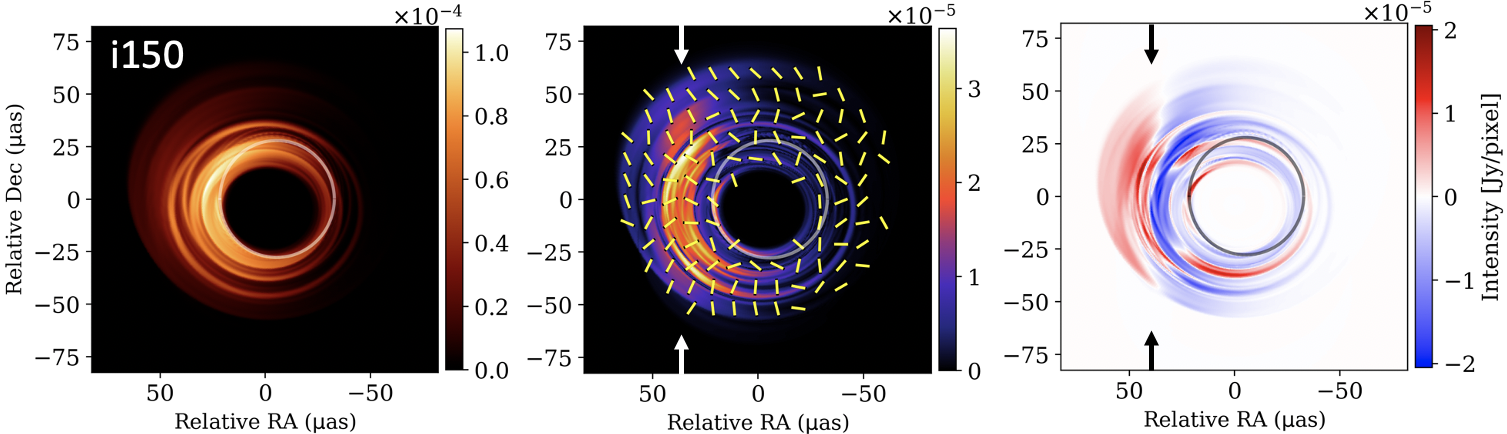}	
	\end{center}
	\caption{Same as figure \ref{i30} but for Model i150. 
	}
	\label{below}
\end{figure*}

\subsection{Comments on possible asymmetric structure}\label{opposite}

\begin{figure*}
	\begin{center}
		\includegraphics[width=16cm]{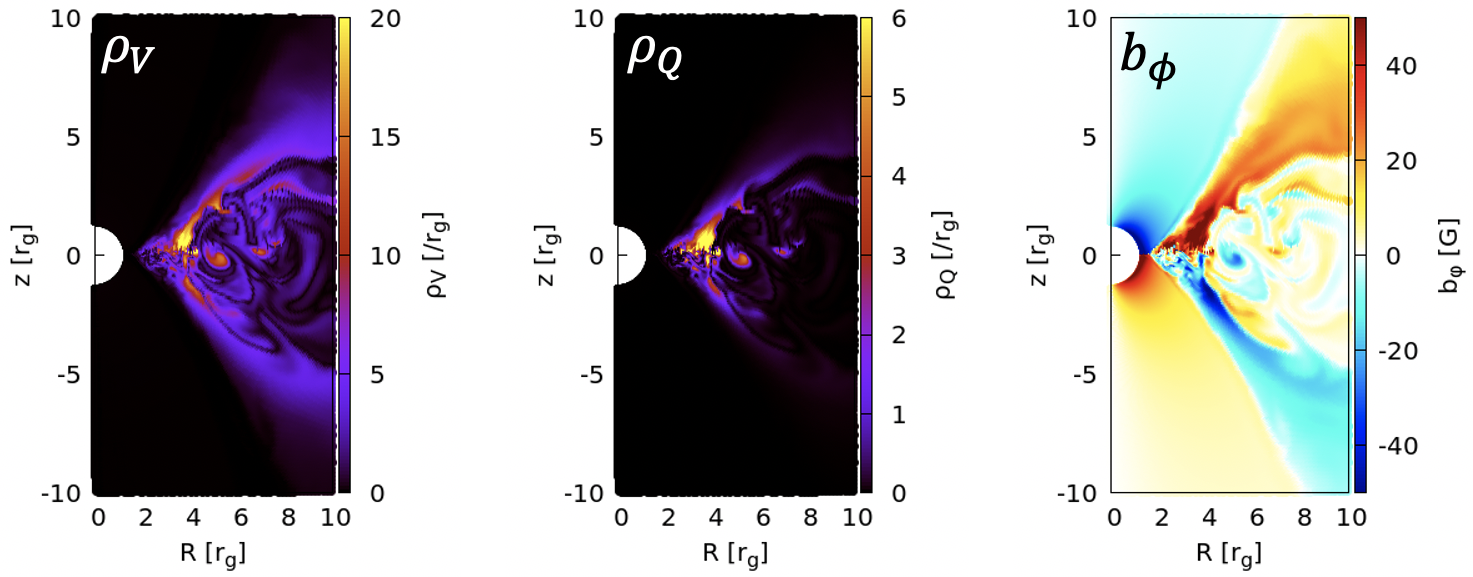}
	\end{center}
	\caption{Left: map of the Faraday conversion coefficient $\rho_Q$ in linear-scale, roughly estimated from electron density, temperature, and magnetic strength. Center: that of the Faraday rotation coefficient $\rho_V$. Right: map of toroidal component of magnetic field in Gauss.
	}
	\label{Fmaps}
\end{figure*}

During the course of the present simulation study, we noticed an asymmetric structural features above and below the equatorial plane in the GRMHD model (in both of gas dynamics and field configurations). 
This means, the images may differ, if we see the disk from the totally opposite direction. We thus repeated the same calculations but with the inclination angle of $i$ = $120\degree$ and $150\degree$, instead of $60\degree$ and $30\degree$, respectively. 
As a result, we found a different feature on the CP image between Model i150 and i30, while no large differences on the images between Model i120 and i60. 
In figure \ref{below} we plot the images for the former i150, the counterpart of those in figure \ref{i30}.

The CP image in the right of figure \ref{below} shows the same tendency as in the right of figures \ref{i30}, not the opposite to them, although the direction of toroidal-dominant magnetic fields are opposite on the two sides of the equatorial plane, as mentioned in subsection \ref{edge-on}.
To survey the apparently inconsistent result, we investigate where the Faraday conversion and rotation occur in the left and central panel of figures \ref{Fmaps}. 
Here we see that these Faraday effects are stronger on the north side than on the south side, as well as the synchrotron emission mentioned above. 
This is because the magnetic field strengths near the black hole are larger in the north at the moment of our plasma models (see the right panel of figures \ref{Fmaps}). 
As a result, observed polarization components reflect the magnetic field configuration in the north side, even when observing from the below (south side of) the equatorial plane, and gives the same sign reversal as in the case observing from the above (north side).
Nevertheless, the right image of figures \ref{below} for Model i150 has a feature that negative components are stronger around the border of sign change (the separatrix of $x_{\rm image} \approx 40~{\rm \mu as}$), different from Model i30, implying the Faraday effects on the south side and the gravitational lensing effect. 
In addition, Model i150 gives CP much higher than Model i30 in total polarization fraction, shown in table \ref{models}, reflecting the stronger Faraday effects in the dense disk to polarization components from the north side. 

Model i120 and i60 give similar values of rotation measure (RM) with the same sign (see table \ref{models}). This also reflects the fact that the Faraday effects mainly occur on the north side. 
By contrast, Model i150 and i30 show RMs with the same sign but with different magnitudes by one order, because of the strong Faraday rotation in the long way from the north to south. 
\citet{Mo17} and \citet{Ri20} demonstrated with M87* in mind that, the Faraday rotation and depolarization mainly occur in the disk rather than the jet, while emission from the foreground jet dominates observed LP flux and RM in their models. 
In our cases, Model i30 is relatively close to their M87* models, where emission on the foreground (on the north side) disk is dominant. Thus the LP flux survives from the Faraday depolarization, and total RM is suppressed. 
In contrast, emission on the background (on the north side) disk is dominated in Model i150 and the LP flux experiences strong Faraday rotation in the disk, as described above, and gives the much larger RM.

\begin{figure*}
	\begin{center}
		\includegraphics[width=16cm]{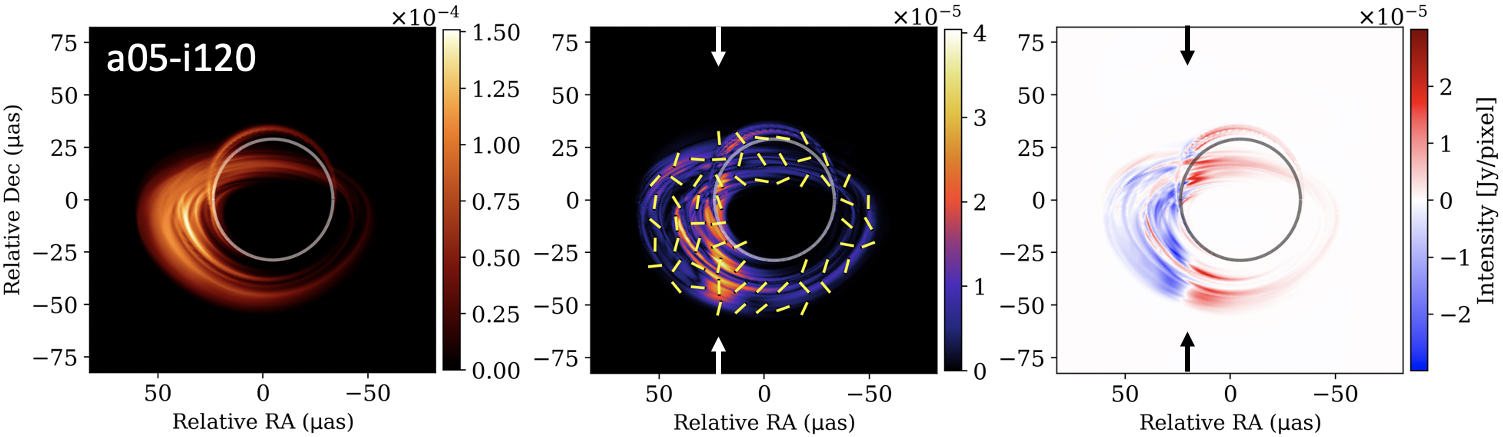}
		\includegraphics[width=16cm]{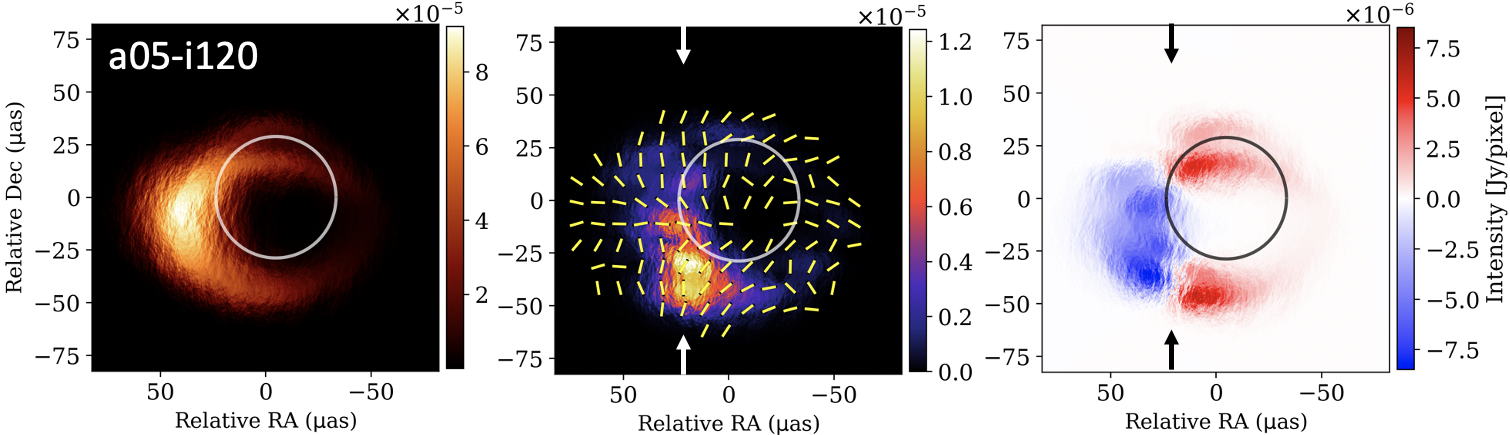}	
	\end{center}
	\caption{Top: the intensity image (left), the LP map (middle), and the CP image (right) for the Model a05-i120. Bottom: same as in the top panel but their scattered images. The positions of the CP separatrix are indicated by the arrows. 
	}
	\label{a05-i120}
\end{figure*}

\subsection{Slow spin case}\label{a05}

As mentioned in section \ref{Intro}, some previous works suggest the fast-spinning black hole and highly magnetized plasma around it (MAD case) to reproduce existing observations of Sgr A*. 
In our study above, we also adjusted models with fast spin of $a=0.9M_{\rm BH}$ and semi-MAD plasma. Here we consider other possibility, a model with slow spin of $a=0.5M_{\rm BH}$ and SANE-like plasma. 

As seen in the bottom row of table \ref{models}, we can reproduce observed total LP and CP flux by adjusting $R_{\rm high}=1$, $i=120\degree$ and $\dot{M} = 3.5\times10^{-8}M_\odot/{\rm yr}$. That is, the electron temperatures are equal to the proton temperature in all over the region. 
In addition, there also exists plasma asymmetry above and below the equatorial plane, as was described in the previous subsection and can be seen through the comparison between Model a05-i60 and  a05-i120 in table \ref{models}. 

The resultant images for a05-i120 are shown in figures \ref{a05-i120}. The total intensity image in the top left panel consists of the beamed disk and the photon ring, similarly to Models i60 and i120 for $a=0.9M_{\rm BH}$ case. The LP image in the top central panel is vertically elongated at $x_{\rm image} \sim 20~{\rm \mu as}$, where the CP image in the top right changes signs of its components. 
The scattered total image in the bottom left panel shows a horizontally-elongated structure because of the extended disk component. In the bottom central and right panels, we can see the sign reversal of CP components and bright LP flux on the CP separatrix, suggesting that these features can also be captured for the SANE-like plasma with the slow spin if there is toroidally-dominant magnetic field.

\begin{figure*}
	\begin{center}
		\includegraphics[width=12cm]{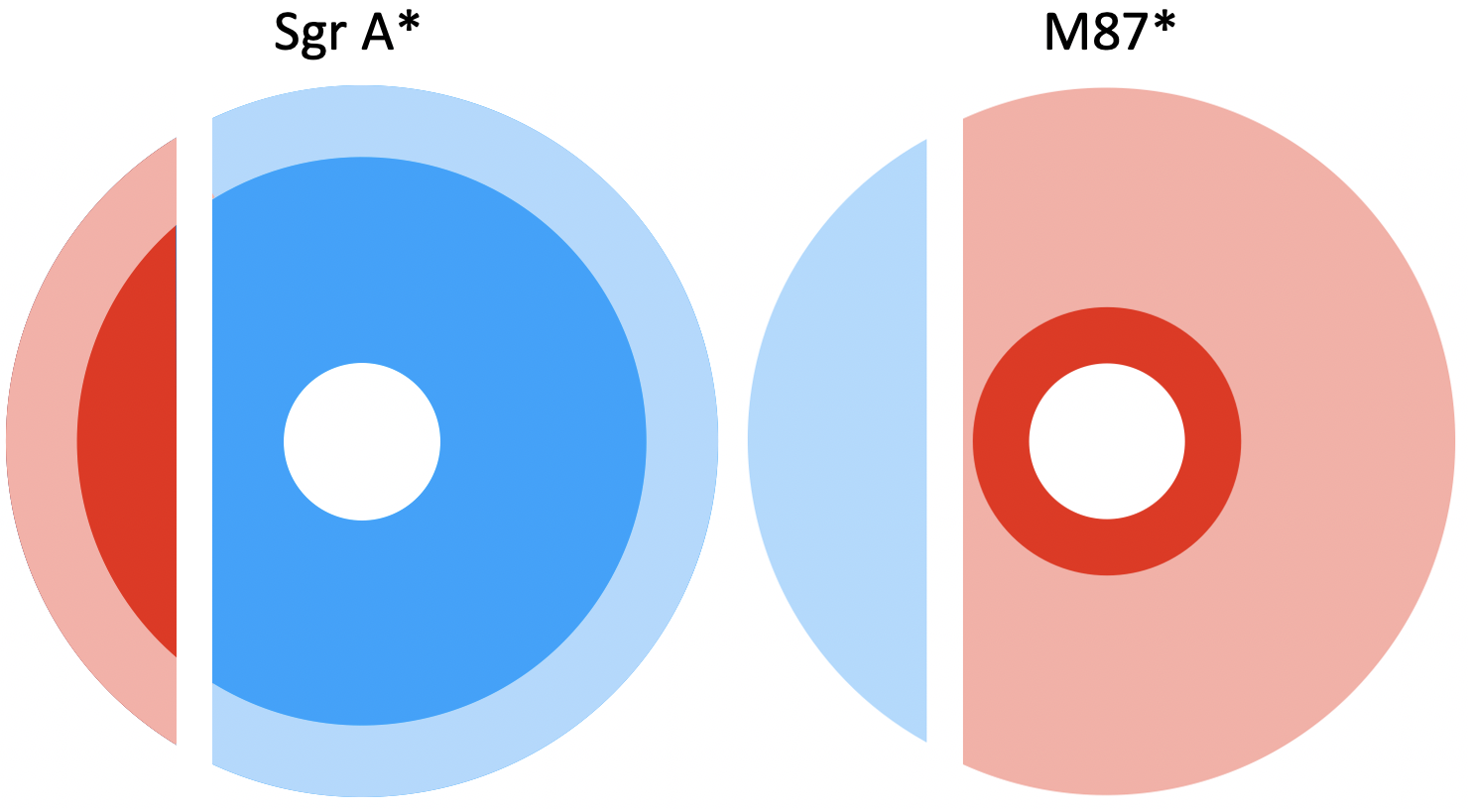}
	\end{center}
	\caption{Schematic view explaining the different CP images obtained for Model i30 in the present study (left panel) and for Model a09R100 in Paper 1 (right panel). The red and blue colors display the region with the circular polarization $V > 0$ and $< 0$ in the observer screen, respectively.  The deep and light colors represent the high and low $|V|$ regions in the screen, repectively.
The red and blue colors are reversed between Sgr A* and M87* because the viewing angles are $i < 90$ deg and $i > 90$ deg in the comparing models, respectively, and because of the difference in the emission region. In the CP image of M87*, the bright region on the image is limited to the right side of the separatrix described in subsubsection \ref{CPamp}, so that the separatrix does not appear in the observed image.	
	}
	\label{SgrAM87}
\end{figure*}

\subsection{Comparison with the case in M87*}

In Paper \Rnum{1}, we obtained the CP ring feature with uniform sign for the fiducial model and other models with low-temperature disk as seen, for example, their figure 2. 
In the context of the CP separatrix described above, here we consider why the CP images for M87* do not show the sign-changing feature and re-interpret them.

In the fiducial model in Paper \Rnum{1} with $a=0.9M_{\rm BH}$, $R_{\rm high}=100$ and $i=160\degree$, both of the synchrotron emission region and the CP amplification region are concentrated to the jet rim at $\sim 5r_{\rm g}$ from the jet axis (cf. their figure 9). 
In addition, the optical depth of the plasma surrounding the BH is less than unity at 230~GHz, so that only the photon ring and its neighborhood shows bright images of Stokes I and V. 

Therefore, the total image and the CP image give compact features around the photon ring ($-5r_{\rm g} \lesssim x \lesssim 5r_{\rm g}$ in the image coordinate), compared with those in this work ($-10r_{\rm g} \lesssim x \lesssim 10r_{\rm g}$ in the image coordinate, here $1~{\rm \mu as} \approx 5r_{\rm g}$ for Sgr A*). 
As a result, the ring feature in the CP image is entirely limited to the right side of the CP separatrix, as seen in a schematic picture of figure \ref{SgrAM87}. 
%In models for SgrA* in this work, by contrast, the plasma around the BH contains the relatively optically-thick region and so the spatially extended structure appear in the Stokes $I$ and $V$ images. 

It should be noted that the features of the image of CP is qualitatively not affected by changing the choose of the magnitude of sigma cut, i.e., the upper limit of the magnetization $\sigma$ ($\equiv B^2/4\pi \rho c^2$) by which the region solved by the polarized GRRT is restricted, as shown in Appendix C in Paper I. 
In this work, we confirmed for Model i30 that the restriction of $\sigma < 1$ does not significantly affect the results and discussions introduced above, and therefore set no sigma-cut as in the fiducial calculations in Paper \Rnum{1}.

There is also a possibility that we can explain the observations of Sgr A* by models with the strong jets like the case of M87*, especially if we observe the jet at low inclination to `hide' the jet (\cite{Mo14,Da18}). 
To verify this possibility, we performed a calculation with $i=20\degree$ and $R_{\rm high}=25$ by using our GRMHD model with $a=0.9M_{\rm BH}$, and obtained an image consistent with the previous works; that is, the image consists of beamed foreground-jet and dim background-jet components while the disk or the photon ring feature are absent.

\subsection{Future prospects}

In the present work, we adopted the axisymmetric GRMHD models to calculate radiative transfer and obtained the polarization images of Sgr A* as a snapshot with the ``fast light'' approximation. 
We have confirmed in the GRMHD simulation data that the flow structure is in a quasi-steady state; that is, the physical quantities do not show large temporal variation but only show rapid, small spatial scale fluctuations occurring on the dynamical timescale. 
Thus we can calculate the observational images by using one snapshot fluid model as a representative one in the quasi-steady state\footnote{After the submission of this paper, we found a paper by \cite{Ri21}, who take up the time-averaged CP images. They show images giving the sign-flipping feature after time-averaged, even in the case where the snapshot image gives turbulent feature.}. 
In addition, since the Keplerian orbital period near the SMBH in Sgr A* ($\sim 10~ {\rm min.}$) is shorter than the observational time of the EHT array ($\sim 10~ {\rm hrs}$), we can expect that the non-axisymmetric features will be smoothed if averaged over the $\phi$-direction by the Keplerian motion. 

Meanwhile, more ``MAD'' models often show much more radical variability and can give variable images in short timescale. 
In future works, we will further research the short-time variabilities and small-scale structure on images for Sgr A* and M87* through time-dependent simulations with 3D GRMHD models, bearing future observations with high resolution in mind. 
There, we will also survey the dependence of the polarimetric features, introduced in this work, on the SANE vs. MAD regime and magnetic field configurations in the disk and the jet.

\section{Conclusion}

We performed full polarimetric radiative transfer simulations based on the semi-MAD models using axisymmetric GRMHD data by \citet{Na18}, and obtained expected linear and circular polarization images in the horizon-scale, bearing EHT observations of Sgr A* in mind. 
For modeling the LLAGN without visible jets, we here assumed a low ratio of the proton temperature and the electron temperature in the disk, and discussed the relationship between the polarimetric features and magnetic field configurations around the SMBH, taking the interstellar scattering effects into account. 
Our results are summarized as follows:
\begin{itemize}
	\item The Stokes $I$ images show the ring-like or the fork-like features for the face-on (Model i30) or the edge-on (Model i90) cases, respectively. The moderate inclination case (Model i60) gives the intermediate feature in between.
	\item In all the three models, the circular polarization images show sign reversals, the so-called ``separatrix'' structure, since the CP amplification through the Faraday effects is proportional to ${\rm cos}\theta_B$ where $\theta_B$ is the angle between the line of sight and the direction of toroidally-dominant magnetic fields. For the same reason (that is, the Faraday rotation is weak for the linear polarization vectors on the separatrix on the images), the linear polarization fluxes are not depolarized and is brightest along the CP  separatrix. These polarimetric features on the linear and circular polarization images can be double evidence of toroidal magnetic fields near the SMBH.
	\item The above features on Stokes $I$ and polarization images still remain, even after undergoing interstellar diffractive and refractive scattering. Further we demonstrated that the angular resolution in present and future EHT observation for Sgr A* at 230~GHz is enough to capture the features described above, and the mitigation of diffractive scattering effects make the situation better. 
	\item When the EHT observations at 345~GHz are available in a near future, we can expect to capture a smaller-scale structure and to directly interpret the polarimetric images in relation to the magnetic field configurations, because all of the synchrotron self-absorption, the Faraday effects, and the interstellar scattering effects are weaker at higher frequency.
\end{itemize}

\bigskip
\section*{Acknowledgement}

The authors wish to acknowledge Masanori Nakamura for provision of GRMHD simulation data sets and stimulating discussions. 
We also thank Jordy Davelaar and Roman Gold for their constructive comments and suggestions.
This work is supported in part by JSPS KAKENHI Grant Number JP20J22986 (YT) and JP18K13594 (TK), JSPS Grant-in-Aid for Scientific Research (A) JP17H01102 (KO), same but for Scientific Research (C) JP17K0583 (SM), JP18K03710 (KO).
This work was supported by MEXT as ``Program for Promoting Researches on the Supercomputer Fugaku'' (Toward a unified view of the universe: from large scale structures to planets, KO, TK) and by Joint Institute for Computational Fundamental Science (JICFuS, KO). 
KA is financially supported in part by grants from the National Science Foundation  (AST-1440254, AST-1614868, AST-2034306). 
Numerical computations were in part carried out on Cray XC50 at Center for Computational Astrophysics, National Astronomical Observatory of Japan. 

\newpage
\onecolumn
\appendix
\setcounter{section}{0} 
\renewcommand{\thesection}{\Alph{section}} 
\setcounter{equation}{0} 
\renewcommand{\theequation}{\Alph{section}\arabic{equation}}
\setcounter{figure}{0} 
\renewcommand{\thefigure}{\Alph{section}\arabic{figure}}
\setcounter{table}{0} 
\renewcommand{\thetable}{\Alph{section}\arabic{table}}
\section{Why is the position of the CP separatrix shifted?}\label{nomotion}

\begin{figure*}
	\begin{center}
		\includegraphics[width=16cm]{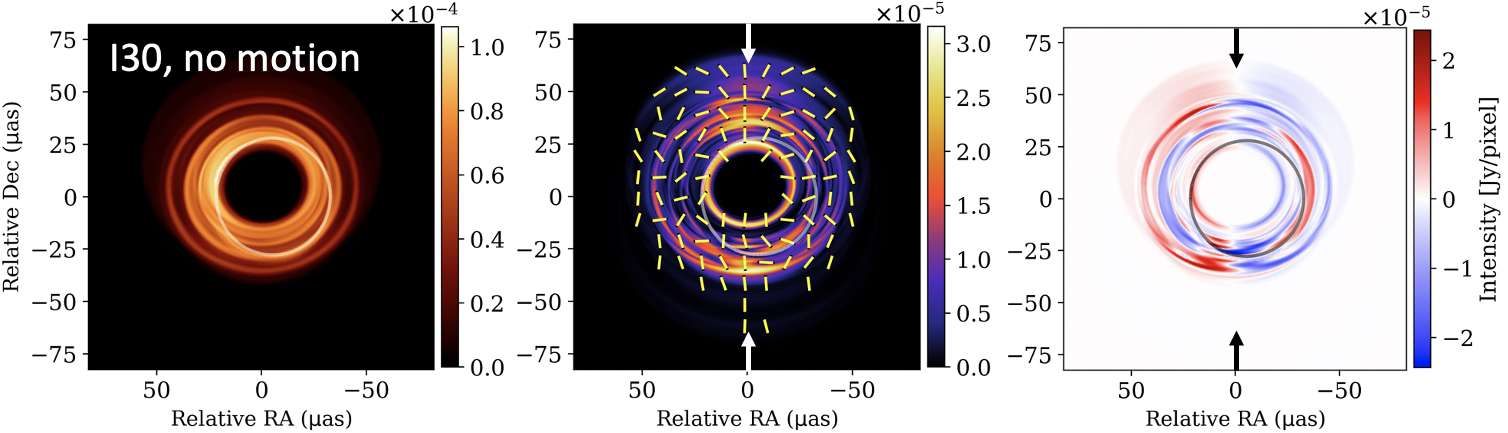}
	\end{center}
	\caption{Same as Fig. \ref{i30} but for the case without plasma bulk motions.
	}
	\label{i30nomotion}
\end{figure*}

In subsubsection \ref{CPamp}, we mentioned that the positional shift of the CP separatrix (to $x_{\rm image} \approx 30~{\rm \mu as}$) is due to the relativistic parallax  effects of plasma bulk motion. 
To demonstrate that this is actually the case, we performed the same simulation but the  case without the plasma bulk motion; i.e., we put the 4-velocity of plasma $u^\mu = (\alpha^{-1},0,0,\alpha^{-1}\omega)$ and thus $u_\mu=(-\alpha,0,0,0)$. 
That is, we take a position of a zero angular momentum observer (\cite{Bar72}). 
Here, $\alpha$ is the lapse function and $\omega$ is the shift vector in $\phi$-direction with minus sign. 

The resultant images are shown in figure \ref{i30nomotion} for the inclination angle of $i=30\degree$. 
As expected, the separatrix in the CP image and the vertically-brightened region in the LP map appear at around $x_{\rm image}\approx 0$. 
This result reflects the toroidal-dominant magnetic fields, because in this case the angles $\theta_B$ are determined only by the magnetic field and light path's vector (line of sight). 
We can thus prove that the asymmetry in the LP and CP images are caused by the special relativistic effects due to plasma bulk motion, and not by the general relativistic effects of the black hole.

\end{document}